\documentclass[aps,prl,twocolumn,showpacs]{revtex4}

\usepackage{amssymb}
\usepackage{amsmath}
\usepackage{graphicx}
\usepackage{natbib}
\usepackage{nicefrac}
\usepackage{multirow}
\usepackage{epstopdf}
\usepackage{float}

\begin{document}

\title{Evidence for Supersolidity in Bulk Solid $^4$He}

\author{Xiao Mi}
\altaffiliation{Present Address: Department of Physics, Princeton University, Princeton, NJ 08544, USA}
\affiliation{Laboratory of Atomic and Solid State Physics and the Cornell Center for Materials Research, Cornell University, Ithaca, New York 14853-2501}
\author{Anna Eyal}
\affiliation{Laboratory of Atomic and Solid State Physics and the Cornell Center for Materials Research, Cornell University, Ithaca, New York 14853-2501}
\author{Artem V. Talanov}
\affiliation{Laboratory of Atomic and Solid State Physics and the Cornell Center for Materials Research, Cornell University, Ithaca, New York 14853-2501}
\author{John D. Reppy}
\affiliation{Laboratory of Atomic and Solid State Physics and the Cornell Center for Materials Research, Cornell University, Ithaca, New York 14853-2501}


\date{July 6, 2014} 

\begin{abstract}
We report low temperature measurements of bulk solid $^4$He in a two-frequency compound torsional oscillator with both annular and open cylinder sample geometries. The oscillators were designed to suppress period shifts arising from all known elastic effects of solid $^4$He. At temperatures below 0.25 K, period shift signals similar to those reported by Kim and Chan [Science {\bf 305}, 1941 (2004)] were observed, albeit two orders smaller in magnitude. A sizable fraction of the observed signals are frequency-independent and consistent with the mass-decoupling expected for supersolid $^4$He. This result is in stark contrast with recent works on Vycor-solid-$^4$He system and suggests that a small supersolid fraction on the order of $1 \times 10^{-4}$ may indeed exist in bulk solid $^4$He.
\end{abstract}

\maketitle
The possible existence of a supersolid, where superflow is supported by the solid phase of $^4$He, was suggested more than forty years ago \cite{Chester_PR1967,Andreev_Defects,Leggett_PRL1970}. Leggett \cite{Leggett_PRL1970} pointed out that a torsional oscillator (TO) containing a solid $^4$He sample would provide an excellent test for the existence of the supersolid state, as the supersolid can be expected to manifest itself in a superfluid-like reduction in the sample moment of inertia,  i.e. a non-classical moment of inertia (NCRI). In 2004, Kim and Chan (KC) made TO measurements for Vycor-solid-$^4$He \cite{KC_Nature2004} and bulk solid $^4$He \cite{KC_Science2004} samples and observed anomalous drops in the resonance periods of the TOs below 0.25 K. The KC results were initially interpreted as evidence for NCRI. However, this interpretation was challenged by the discovery, by Day and Beamish (DB) \cite{DB_Nature2004}, of a temperature-dependent anomaly in the shear modulus, $\mu$, of solid $^4$He, occurring over the same temperature range as the KC TO signals as well as sharing the same dependence on velocity/strain and $^3$He impurity level. A re-examination by two separate groups \cite{Mi_PRL2012,Kim_PRL2012} of the KC discovery in Vycor-solid-$^4$He were recently performed. The conclusion drawn from these experiments is that the original KC period shift observation in Vycor-solid-$^4$He arose from the increasing $\mu$ of a thin layer of bulk solid $^4$He in the TO cells and was unrelated to supersolidity. On the other hand, the possible existence of supersolidity in bulk solid $^4$He still remains an open theoretical and experimental question \cite{Anderson_Arxiv2013}.

Multiple-frequency TOs provide an effective method for identifying the origins of the anomalous period shifts below 0.25 K for samples of solid $^4$He. In the case of a supersolid NCRI, the fractional period shift (FPS), defined as the anomalous period drop normalized by the mass-loading sensitivity, is expected to be independent of the TO frequency. The FPS is determined by $\Delta P / \Delta P_{\text{F}}$, where $\Delta P$ is the magnitude of the anomalous period drop and $\Delta P_{\text{F}}$ is the increase in the period of the TO upon freezing of the sample. Should the observed period drop arise from the shear-stiffening of solid $^4$He, the FPS would assume different values at different frequencies. The FPS values obtained at different frequencies allow the decomposition of the observed signals into frequency-dependent and frequency-independent contributions. The existence of a finite frequency-independent contribution to the FPS could then indicate the presence of supersolid NCRI.

The shear-stiffening of solid $^4$He \cite{DB_Nature2004} can alter the TO period in multiple ways, including period shifts arising from the acceleration of solid $^4$He \cite{Reppy_JLTP_Review2012}, stiffening of the solid $^4$He in the torsion rods \cite{Beamish_TorsionRodEffect}, the counter-stress of solid on the cell wall \cite{Maris_Effect} and through the dissipative dynamics of the solid \cite{Graf_Glassy_DoubleTO}. It is critical in the design of a double-frequency TO that the significant elastic effects should be understood and reduced to a minimum in order to facilitate the analysis of the observed FPS. Such conditions have not been met in the earlier double-frequency experiments on bulk solid $^4$He \cite{Kojima_PRL_DoubleTO,Cowen_DoubleTO_2013}. In these experiments, the FPS signals contained significant contributions from both the acceleration of solid $^4$He and the stiffening of solid $^4$He inside the torsion rods. In Ref.~\cite{Mi_JLTP2014}, we reported the preliminary result from a double-frequency TO that was designed to be chiefly sensitive to elastic effect due to acceleration of solid $^4$He. In this experiment, a frequency-independent contribution to the FPS was observed that was equivalent to an inertial-mass-decoupling of the cylindrical sample corresponding to a NCRI/supersolid fraction of $1.2 \times 10^{-4}$.

\begin{figure}
\centering
\includegraphics[width=\columnwidth]{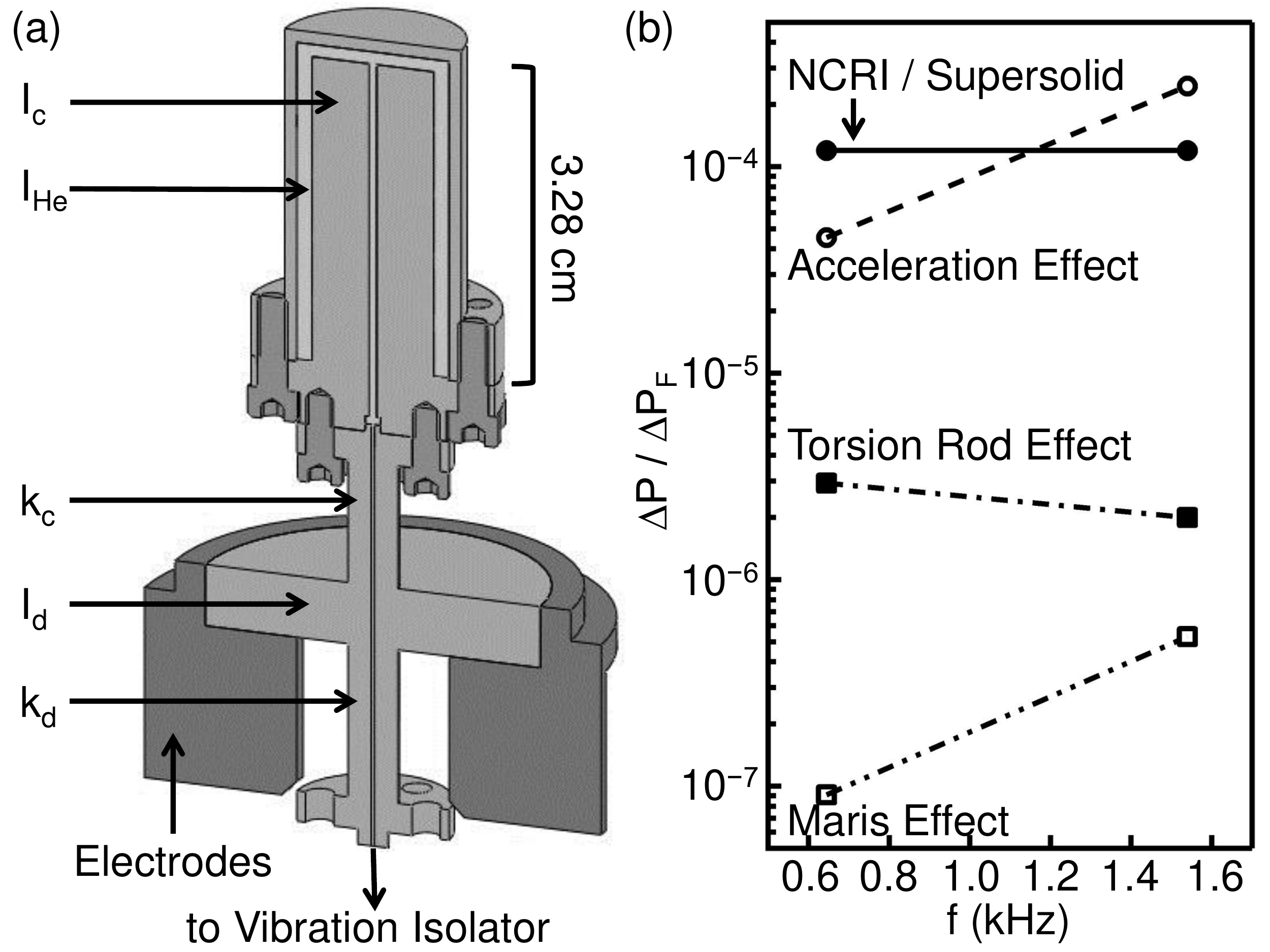}
\caption{(a) To-scale drawing of the double-frequency TO. The electrodes for drive/detection of TO motion are constructed from Mg. The rest of the TO is made of annealed Al 6061. (b) Summary of FPS from various effects calculated by varying $\mu$ from $1.5 \times 10^8$ dyn$\,$cm$^{-2}$ to $3.0 \times 10^8$ dyn$\,$cm$^{-2}$, at each frequency of the TO. Value for the acceleration effect is based on FEM computation; the others are analytical estimates.}
\label{fig:1}
\end{figure}

The encouraging result from our cylindrical cell \cite{Mi_JLTP2014} prompted a further investigation into the possibility of supersolidity in bulk solid $^4$He. In a new experiment, an annular sample geometry was employed which allows a further reduction in the elastic contributions to the FPS signals. Fig.~\ref{fig:1} (a) shows the cross-section of the double-frequency TO constructed from annealed Al 6061. The moments of inertia of the cell holding solid $^4$He and the dummy oscillator are $I_\text{c} = 26.0$ g$\,$cm$^2$ and $I_\text{d} = 58.7$ g$\,$cm$^2$. The corresponding torsion rod constants are $k_\text{c} = 1.69 \times 10^9$ dyn$\,$cm and $k_\text{d} = 1.35 \times 10^9$ dyn$\,$cm. The torsion rods have inner fill-lines with radius, $r_\text{fill} = 0.017$ cm and outer radius $r_\text{rod} = 0.256$ cm, for a ratio of $r_\text{fill} / r_\text{rod} = 0.067$. This small ratio greatly reduces the elastic effect from solid $^4$He in the fill-lines \cite{Beamish_TorsionRodEffect}. The sample is largely annular and has a total solid moment of inertia $I_\text{He} = 0.249$ g$\,$cm$^2$. A notable improvement in this TO over our previous efforts \cite{Mi_PRL2012,Mi_JLTP2014} results from the mounting of the double oscillator on an additional vibration isolator (VI), which is itself a TO with a massive moment of inertia $I_\text{v} = 547$ g$\,$cm$^2$ and torsion constant $k_\text{v} = 1.03 \times 10^9$ dyn$\,$cm. The VI is in turn mounted to a Cu block that is thermally anchored to the mixing chamber of the dilution refrigerator. The addition of the VI has improved the signal-to-noise ratio by a factor of ten, allowing detection of signals as small as $\Delta P = 0.01$ ns, or FPSs on the order of $5 \times 10^{-6}$.

In a Supplement to this Letter, we provide detailed calculations of the known elastic effects of solid $^4$He for this TO, using both an analytical approach and a finite element method (FEM). Based on these calculations, the estimated FPS values for each elastic effect at the two resonance frequencies are shown in Fig.~\ref{fig:1}(b). It is clear that, for this annular cell, the only elastic effect on the order of $10^{-4}$ arises from the acceleration of solid $^4$He, with a FPS proportional to $f^2$, the square of TO frequency.

\begin{figure}
\centering
\includegraphics[width=\columnwidth]{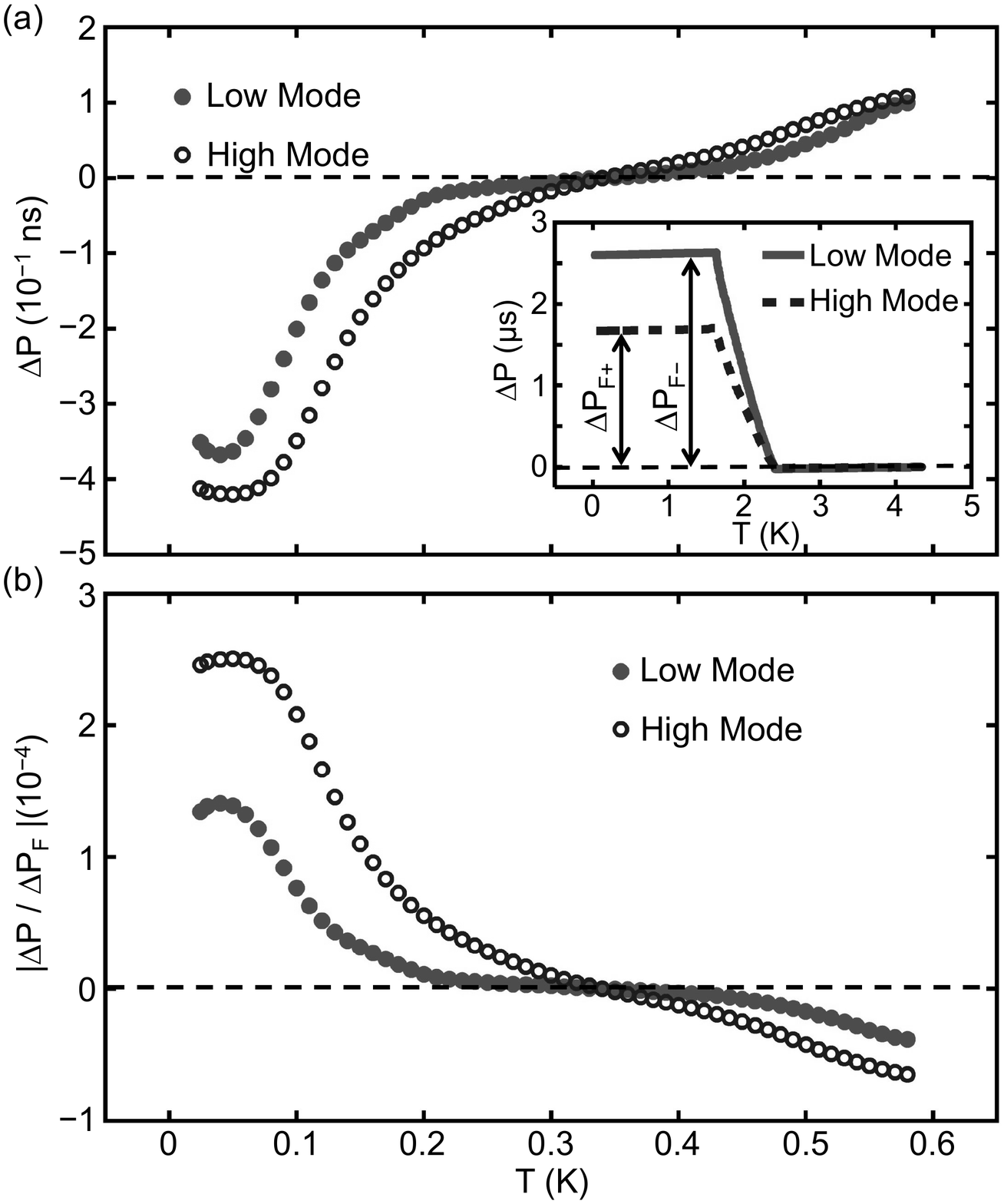}
\caption{(a) Period shifts $\Delta P$ at each resonance mode, defined as $\Delta P (T) = P (T) - P(T = 0.34 \text{ K})$ where $P(T)$ is the period of the oscillator after correction for the temperature dependent empty cell background. Inset: Period shifts upon freezing of the sample. Mass-loading sensitivities of $\Delta P_{\text{F}-} = 2.75$ $\mu$s and $\Delta P_{\text{F}+} = 1.69$ are extracted. (b) The magnitude of the FPS is shown as a function of temperature for each mode.}
\label{fig:2}
\end{figure}

We refer the reader to Ref.~\cite{Mi_PRL2012} for a complete description of the drive/detection scheme of the double-frequency TO. The two frequency modes are excited simultaneously with identical maximum rim velocity of 6 $\mu$ms$^{-1}$ for each mode. We have also driven each mode individually to avoid any nonlinear mode-coupling effects and found the results presented in this Letter to be independent of the method of operation. At $T = 0.5$ K, the resonance periods of the empty TO are $P_- = 1.548$ ms for the low frequency $(-)$ mode and $P_+ = 0.648$ ms for the high frequency $(+)$ mode, with corresponding frequencies of $f_- = 646.0$ Hz and $f_+ = 1543.0$ Hz. Since $I_\text{v} \gg I_\text{c}$ and $I_\text{v} \gg I_\text{d}$, the influence of the VI on the resonance periods of the TO is small and the two resonance periods are well approximated by $P_\pm = 2 \pi \left[ \tfrac{I_\text{c} \left( k_\text{c} +  k_\text{d} \right) +I_\text{d} k_\text{c}}{2 I_\text{c} I_\text{d}} \left( 1 \pm \sqrt{1 - \tfrac{4 I_\text{c} I_\text{d} k_\text{c} k_\text{d}}{(I_\text{c} \left( k_\text{c} +  k_\text{d} \right) +I_\text{d} k_\text{c})^2}} \right) \right]^{- 1/2}$. 

A polycrystalline sample is formed from commercial $^4$He gas having a nominal 0.3 ppm $^3$He impurity level by the blocked-capillary method. The inset to Fig.~\ref{fig:2}(a) shows the period shift data for the two resonance modes as the sample is frozen. The total period shifts for the two modes, between the liquid and the solid phase, are $\Delta P_{\text{F}-} = 2.75$ $\mu$s and $\Delta P_{\text{F}+} = 1.69$ $\mu$s. These shifts determine the mass-loading sensitivities for the two modes. Based on a temperature of 1.63 K at which freezing ceases, the final sample pressure is about 28 bar.

In Fig.~\ref{fig:2}(a), we present data for the period shift $\Delta P (T)$ at each mode as a function of temperature from T = 0.02 K to T = 0.58 K. $\Delta P (T)$ is defined taking the cell period at $T = 0.34$ K as the reference value, i.e., $\Delta P (T) = P (T) - P(T = 0.34 \text{ K})$. We observe clear anomalous period drops below 0.2 K where the $\Delta P$ decreases rapidly. In the intermediate regime, $0.2 \text{ K} < T < 0.4 \text{ K}$, $\Delta P$ shows little variation over temperature. At $T > 0.4 \text{ K}$, a moderate increase in $\Delta P$ is observed which continues to higher temperatures. As a next step, we normalize $\Delta P$ at each mode by its mass-loading sensitivity $\Delta P_\text{F}$ to obtain the FPS. The results are plotted in Fig.~\ref{fig:2}(b). If the observed signals were attributed entirely to supersolidity, we would expect the FPS to be frequency-independent and depend only on temperature. The two curves in Fig.~\ref{fig:2}(b) would coincide in this case. This is not the case, with the difference attributable to the elastic effect arising chiefly from the acceleration of the $^4$He solid. If the signals arise entirely from the elastic effects of solid $^4$He, we would expect the FPS at the two modes to follow the relation $\Delta P_+ / \Delta P_{\text{F}+} = (f_+ / f_-)^2 (\Delta P_- / \Delta P_{\text{F}-}) \approx 5.7 (\Delta P_- / \Delta P_{\text{F}-})$, which is also not the case. Therefore, we shall analyze the FPS signals as composites consisting of two components, a frequency-independent component (supersolid fraction) and a component proportional to the square of the frequency arising from dynamic elastic effects.

\begin{figure}
\centering
\includegraphics[width=\columnwidth]{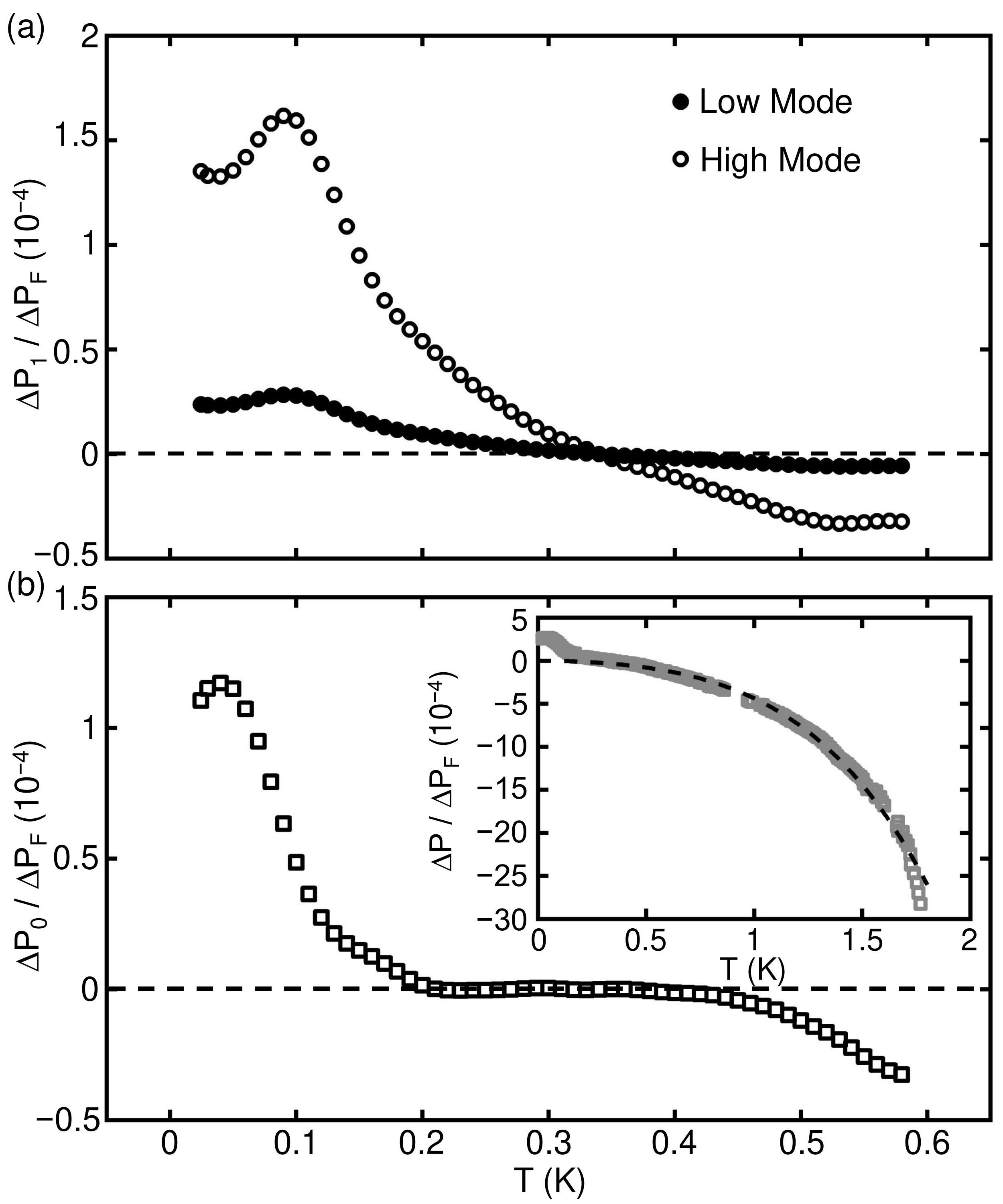}
\caption{(a) Elastic, frequency-dependent, contributions to the FPS, $\Delta P_1 / \Delta P_F$, at both modes. (b) Supersolid, frequency-independent, contribution to the FPS, $\Delta P_0 / \Delta P_\text{F}$, identical for both modes. Inset: Total FPS up to melting temperature $\Delta P / \Delta P_\text{F}$ for a different sample in a single-mode TO similar in design to the cell used in this experiment. The dashed line, through the data in the inset, is a fit to the functional form of the temperature-dependent variation in solid $^4$He pressure \cite{Pressure_GlassyHe}, $A T^4 + B T^2$.}
\label{fig:3}
\end{figure}

In decomposing the FPS into individual contributions at each temperature $T$, we proceed as follows: For each mode, the FPS can be written as $\Delta P_\pm (T) / \Delta P_{\text{F}\pm} = \Delta P_0 (T) / \Delta P_{\text{F}\pm} + \Delta P_{1\pm} (T) / \Delta P_{\text{F}\pm}$, where $\Delta P_0 (T)$ is a temperature-dependent constant and $\Delta P_0 (T) / \Delta P_{\text{F}\pm}$ is the temperature-dependent supersolid contribution to the FPS. The elastic contribution to the FPS is $\Delta P_{1\pm} (T) / \Delta P_{\text{F}\pm} = C_0 (T) f_\pm^2$, where $C_0 (T)$ is a temperature-dependent constant. In Fig.~\ref{fig:3}(a), we plot the elastic contributions to the FPS as a function of temperature. It can be seen that the elastic contribution continues to change above $T$ = 0.2 K. This feature is consistent with the continuing change in $\mu$ above 0.2 K, and is seen in previous TO experiments that are dominated by elastic effects \cite{Mi_PRL2012,Mi_Triple}. Based on a FEM computation described in the Supplement, we deduce that a 50$\%$ increase in $\mu$ with a base value of $1.5 \times 10^8$ dyn$\,$cm$^{-2}$ would produce the observed elastic contributions to the FPS. This is in good agreement with previous measurements of shear-stiffening of solid $^4$He, which have reported changes up to 80$\%$ \cite{Paalanen_Shear_1981}.

More interesting is the frequency-independent term or supersolid contribution to the FPS shown in Fig.~\ref{fig:3}(b). Below 0.2 K, this contribution increases to a maximum of $1.2 \times 10^{-4}$. As the temperature is raised, the frequency-independent contribution declines and becomes essentially constant between 0.2 K and 0.4 K, suggesting 0.2 K as the approximate supersolid transition temperature for this sample. This behavior is consistent with a zero supersolid contribution for $T > 0.2$ K and contrasts with the elastic contribution, which changes continuously over this temperature range. Above 0.2 K, the frequency-independent contribution begins to decline below zero and becomes increasingly negative with increasing temperature. We believe that this trend is explained, in part, by a pressure-induced expansion of the cell. The pressure of constant-volume solid $^4$He sample \cite{Pressure_GlassyHe} increases as $p(T) = A T^4 + B T^2 + p_0$. As $p(T)$ increases with increasing temperature, the cylindrical walls of the cell are expanded, causing the moment of inertia of the cell, $I_\text{c}$, to increase. The pressure effect becomes much more visible at higher temperatures, as illustrated in the inset of Fig.~\ref{fig:3}(b), where data from an earlier single-mode TO are shown for temperatures up to 1.8 K. 

\begin{figure}
\centering
\includegraphics[width=\columnwidth]{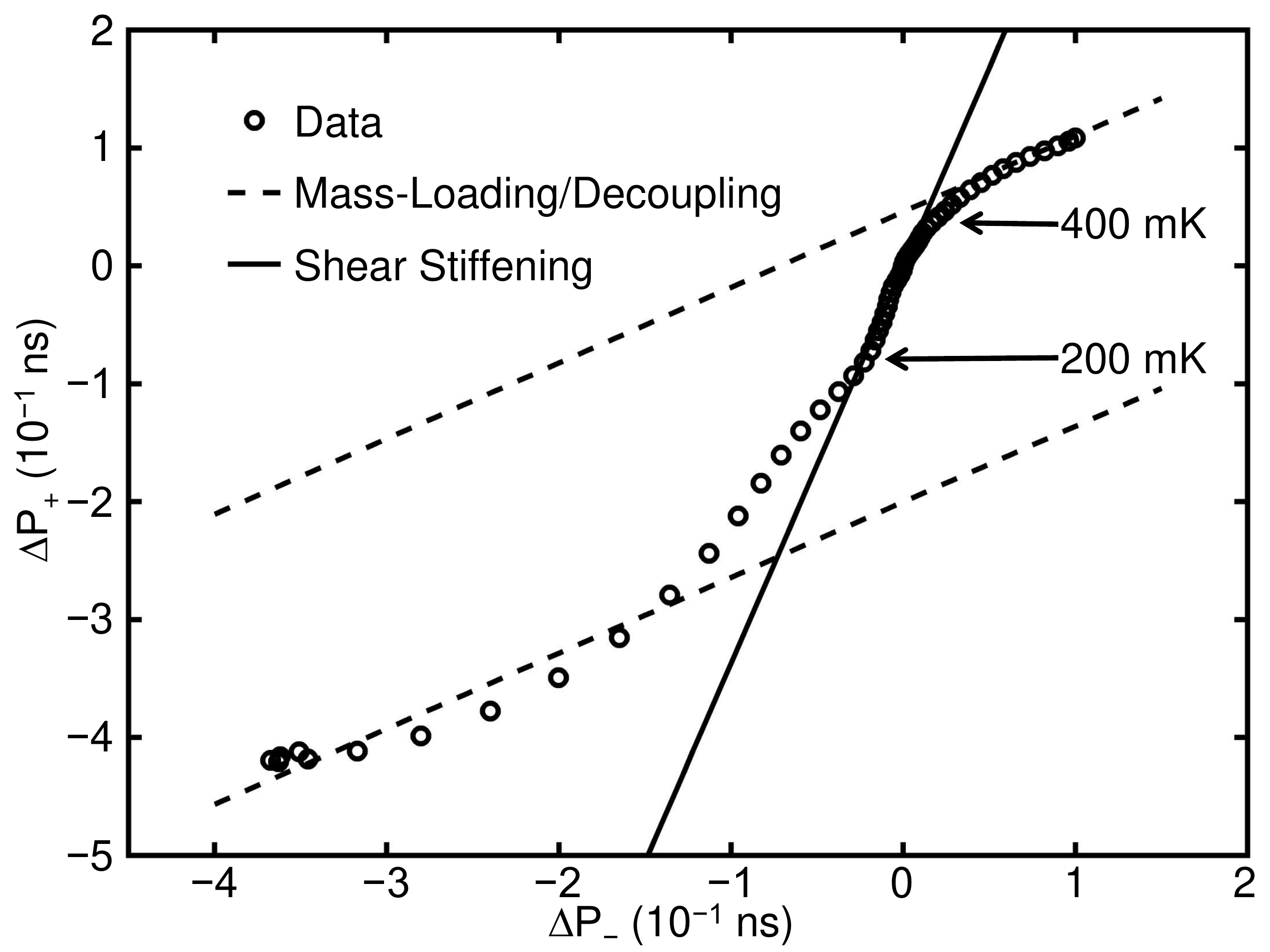}
\caption{$\Delta P_+$ against $\Delta P_-$ using temperature-dependent values from Fig.~\ref{fig:2}(a). Slopes corresponding to a mass-loading/decoupling scenario and a solid $^4$He shear-stiffening scenario are included.}
\label{fig:4}
\end{figure}

An instructive way to visualize the data is to plot the period shifts $\Delta P_\pm$ at different temperatures against each other, as shown in Fig.~\ref{fig:4}. In the scenario of mass-loading/decoupling, the data would follow a slope equal to the ratio of mass-loading sensitivities, $\Delta P_{\text{F}+} / \Delta P_{\text{F}-}$. Should the signals arise from shear-stiffening of solid $^4$He, the slope would be equal to $(f_+/f_-)^2 \Delta P_{\text{F}+} / \Delta P_{\text{F}-}$. We see that above 0.4 K, the data follow the mass-loading slope, consistent with pressure-driven expansion of the cell. Between 0.2 K and 0.4 K, the slope of the data is consistent with the expectations based on shear-stiffening of solid $^4$He. Below $T = 0.2$ K, the slope of the data returns to a value close to that of mass-decoupling, indicating that the majority of the observed signals, in this temperature range, is due to the presence of supersolidity.

\begin{figure}
\centering
\includegraphics[width=\columnwidth]{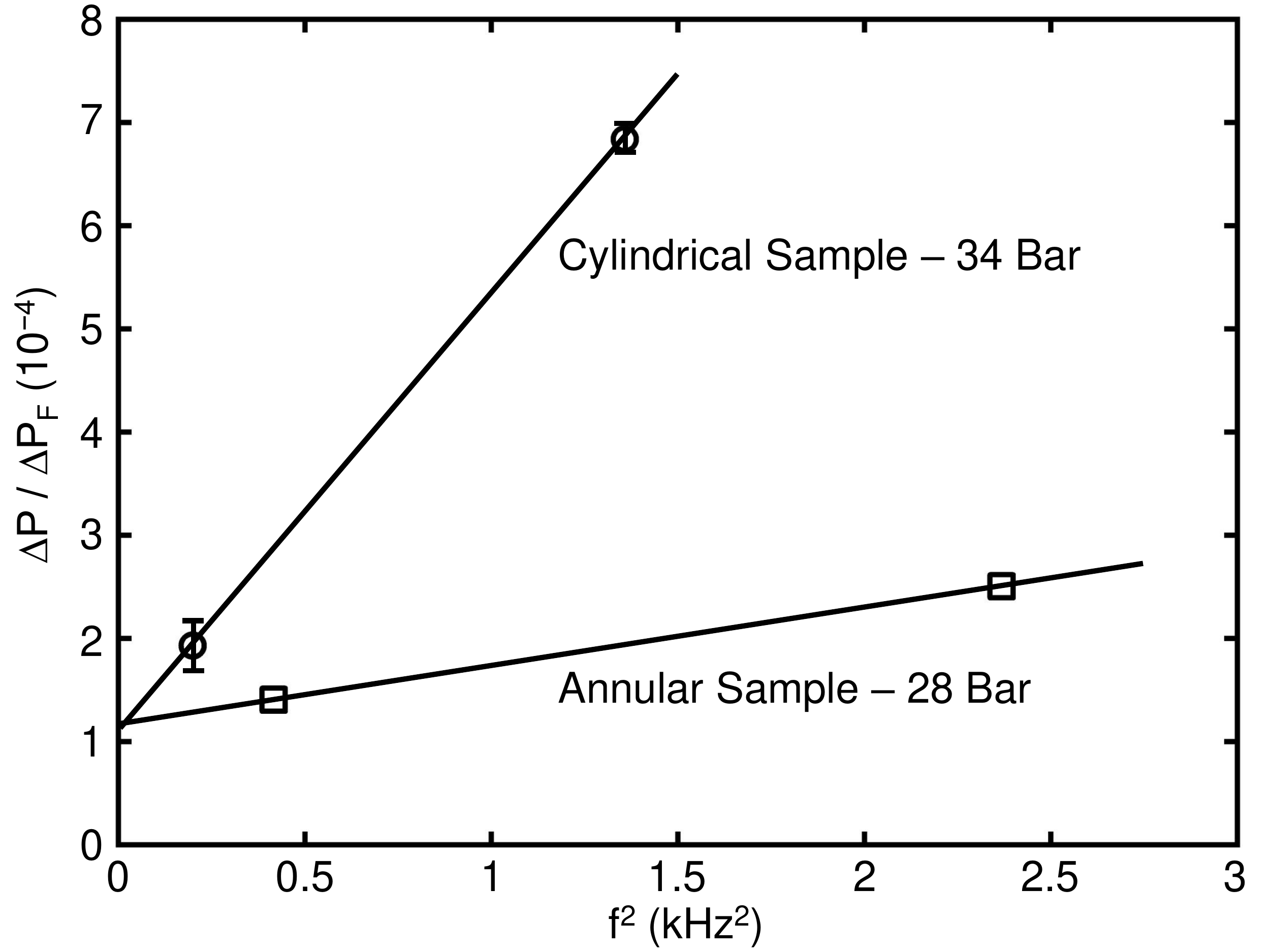}
\caption{FPS at $T = 0.02$ K for two double-frequency TO's plotted against $f^2$. Error bars are larger in the cylindrical cell due to greater noise level, and are smaller than the size of the symbols for the annular cell.}
\label{fig:5}
\end{figure}

In Fig.~\ref{fig:5}, we provide a comparison of the results from this annular cell with those from our previous cylindrical cell \cite{Mi_JLTP2014}. For each cell, the FPS at 0.02 K for the low and high modes are plotted against the squares of their respective frequencies, $f^2$. We see that the frequency dependence of the FPS is much weaker in the annular cell, as expected due to the greater suppression of the elastic effect arising from inertial acceleration of solid $^4$He. The zero-frequency $y$-intercept gives a non-zero supersolid fraction for both TOs. Despite the complete independence of the two experiments and their difference in geometry, the supersolid fraction is identically $(1.2 \pm 0.1) \times 10^{-4} $. 

Since the KC discovery in 2004, TO experiments performed on bulk solid $^4$He have reported FPS values ranging from $4 \times 10^{-4}$ to 0.2 \cite{KC_Science2004,Kojima_PRL_DoubleTO,Cowen_DoubleTO_2013,Mi_Triple,Rittner_Disorder,Clark_PRL2007,Shirahama_JLTP148,Kubota_JLTP158,Rittner_UpperLimit,Kubota_PRL2008,Hunt_Science2009,Kojima_DoubleTO_2010,Reppy_Plastic_Deformation,Pratt_Science2011,DYKim_2011_PRB,Golov_PRL107,Fefferman_2012_TO}, with the exceptions of an early spherical TO experiment by Bishop et al \cite{Bishop_TO1981}, which saw no supersolid fraction above $1 \times 10^{-4}$, and the rigid TO experiments performed at Penn State \cite{Long_PathLength_TO,Moses_NewResult}, which have placed an upper bound of $4 \times 10^{-6}$ on the supersolid fraction.  It is pointed out in \cite{Long_PathLength_TO} that the mass flow observed by Ray and Hallock \cite{Hallock} at low solid pressures, if limited to a flow velocity of 10 $\mu$m/sec, could lead to FPS on the order $1 \times 10^{-4}$, similar to what is observed in their long path-length experiments and in the results we report here.
In retrospect, many of the past experiments are heavily influenced by different elastic effects of solid $^4$He, leading to recent reports of the non-existence of supersolidity in $^4$He \cite{Physics_News_NoSupersolid}. However, the shear modulus anomaly of solid $^4$He may very well coexist, or possibly be correlated, with supersolidity \cite{Anderson_Arxiv2013}. The small supersolid fraction is simply obscured by huge elastic effects for these experiments. Indeed, the value of the possible supersolid fraction reported here ranks among the smallest seen prior to this work. Therefore, our observation is consistent with the tortuous history of solid $^4$He and indicates that a small superflow can persist underneath the giant elastic anomaly. We conclude by noting that the magnitude of the supersolid fraction is consistent with past theoretical predictions \cite{Leggett_PRL1970}.

\begin{acknowledgements}
We acknowledge useful and encouraging discussions with P. W. Anderson, W. F. Brinkman and D. A. Huse. This work was supported by the National Science Foundation through Grant DMR-060586 and CCMR Grant DMR-0520404, and partially funded by the New England Foundation, Technion.
\end{acknowledgements}

\bibliographystyle{apsrev}
\bibliography{references}

\end{document}


\title{Supplementary Material for ``Evidence for Supersolidity in Bulk Solid $^4$He''}

\author{Xiao Mi}
\altaffiliation{Present Address: Department of Physics, Princeton University, Princeton, NJ 08544, USA}
\affiliation{Laboratory of Atomic and Solid State Physics and the Cornell Center for Materials Research, Cornell University, Ithaca, New York 14853-2501}
\author{Anna Eyal}
\affiliation{Laboratory of Atomic and Solid State Physics and the Cornell Center for Materials Research, Cornell University, Ithaca, New York 14853-2501}
\author{Artem V. Talanov}
\affiliation{Laboratory of Atomic and Solid State Physics and the Cornell Center for Materials Research, Cornell University, Ithaca, New York 14853-2501}
\author{John D. Reppy}
\affiliation{Laboratory of Atomic and Solid State Physics and the Cornell Center for Materials Research, Cornell University, Ithaca, New York 14853-2501}

\date{July 7, 2014}

\maketitle

In this Supplement, we provide a more detailed discussion of the effects of shear modulus changes in solid $^4$He on the resonance periods of the torsional oscillator (TO) employed in the Letter. We approach the problem with an analytical calculation and a numerical simulation using Finite Element Method (FEM). It is assumed that the absolute value of the shear modulus of solid $^4$He is independent of frequency in the frequency range of our TO. This is justified by the measurements of Day et al. \cite{DB_Nature2004} who observed a mere 1$\%$ variation in the shear modulus of solid $^4$He at frequencies between 200 Hz and 2000 Hz.

\section{Analytical Calculation}

\subsection{Acceleration Effect}

We first calculate the effect arising from the acceleration field of solid $^4$He. The majority of the solid $^4$He sample is confined in a long annular channel with inner radius $r_\text{i} = 0.635$ cm, outer radius $r_\text{o} = 0.794$ cm and height $L = 3.28$ cm. The remaining solid $^4$He sample is confined at the top of the sample volume and in a thin cylindrical space with radius $r_\text{c} = r_\text{i} = 0.635$ cm and height $H = 0.127$ cm.

For the annular part of the sample, the amplitude of the displacement field, $\vec{u}$, as a function of radius from the symmetry axis of the sample $r$ is calculated to be \cite{Reppy_JLTP_Review2012}
\begin{equation}
\vec{u}(r) = \frac{r_\text{m} \theta_{0} \cos \left( (r - r_\text{m}) \omega \sqrt{\rho / \mu} \right)}{ \cos \left( \tfrac{1}{2} \Delta r \omega \sqrt{\rho / \mu} \right) } \vec{e}_{\theta}
\end{equation}
where $\Delta r = r_\text{o} - r_\text{i} = 0.159$ cm is the width of the annulus, $r_\text{m} = \tfrac{1}{2} (r_\text{o} + r_\text{i}) = 0.715$ cm is the mean radius of the annulus, $\omega = 2 \pi f$ where $f$ is TO frequency which takes on values of $f_-$ and $f_+$ at the two resonance modes, $\rho$ and $\mu$ are the density and shear modulus of solid $^4$He, $\theta_0$ is the maximum angular displacement in radians of the oscillating TO and $\vec{e}_{\theta}$ is the azimuthal direction defined with $z$-axis being the symmetry axis of the sample. In this estimate, we are neglecting the finite length of the annulus since it is much greater than the radius of the annulus, $L \gg r_\text{m}$. The first internal radial sound mode of solid $^4$He in this geometry occurs at a frequency of $f_s = \tfrac{1}{2 \Delta r} \sqrt{\mu / \rho} = 86.3$ kHz, nearly two orders of magnitude higher than the TO frequencies. We are therefore justified in expanding $\vec{u}(r)$ in terms of $\omega$ and keeping only the two lowest order terms.

To convert the displacement field into a back-action torque $\tau_\text{He}$ on the TO, we integrate the $\theta$ component of the displacement field according to $\tau_\text{He} = \int_{r_\text{i}}^{r_\text{o}} dr \rho \omega^2 u_\theta (r)$ \cite{Reppy_JLTP_Review2012,Graf_Glassy_DoubleTO}. The effective moment of inertia of solid $^4$He is then
\begin{equation}
I_\text{eff} = \tau_\text{He} / (\theta_0 \omega^2) \approx I_\text{He} \left( 1 + \frac{1}{2} (\Delta r)^2 \omega^2 \frac{\rho}{\mu} \right)
\end{equation}
where $I_\text{He} = \tfrac{1}{2} \rho L \pi (r_\text{o}^4 - r_\text{i}^4) = 0.242$ g$\,$cm$^{2}$ is the moment of inertia of the annular part of the $^4$He sample. An increase in $^4$He shear modulus decreases $I_\text{eff}$ by an amount $\Delta I_\text{eff} \propto \omega^2$, which leads to a fractional period shift (FPS), $\Delta P / \Delta P_\text{F} = \Delta I_\text{eff} / I_\text{eff}$. Therefore, the observed FPS is proportional to $f^2$. The cylindrical part of the sample is treated in analogous manner. The effective moment of inertia has a similar form \cite{Reppy_JLTP_Review2012}:
\begin{equation}
I_\text{eff} \approx I_\text{He} \left( 1 + \frac{1}{2} H^2 \omega^2 \frac{\rho}{\mu} \right)
\end{equation}
The moment of inertia of this part of the sample is $I_\text{He} = \frac{1}{2} \rho H \pi r_\text{c}^4 = 0.0065$ g$\,$cm$^2$.

To estimate the magnitude of FPS at each resonance mode due to shear-stiffening of solid $^4$He, we vary $\mu$ from $1.5 \times 10^8$ dyn$\,$cm$^{-2}$ to $3 \times 10^8$ dyn$\,$cm$^{-2}$ and compute the fractional change in the total effective moment of inertia which is the sum of expressions in eqns (2) and (3). The result is a FPS of $1.32 \times 10^{-4}$ for the high frequency mode and $0.23 \times 10^{-4}$ for the low frequency mode, both being very close to the elastic contributions to our observed FPS. We note that the 100$\%$ increase in $^4$He shear modulus may seem unlikely. This is because we have ignored the finite shear modulus of the aluminum alloy constituting the TO. The next effect we discuss takes the elasticity of the cylindrical wall of the cell into account. In the FEM computation presented in the next section, the finite shear modulus of the entire TO is taken into account which amplifies the FPS calculated here, although the FPS remains proportional to $f^2$.

\subsection{Twisting of Cell Walls}
In Ref. \cite{Reppy_JLTP_Review2012}, we discussed a correction to the TO periods due to the twisting of the cell walls. This effect arises from the fact that the top of the cell undergoes a slightly larger angular displacement than the bottom of the cell, due to the elastic displacement of the cylindrical cell wall. The size of this effect is approximated by changes in the effective moment of inertia $I_\text{E}$ of the cell itself, which is given by
\begin{equation}
I_\text{E} \approx I_\text{U} \left( 1 + \frac{1}{3} U^2 \omega^2  \frac{\bar{\rho}}{\bar{\mu}} \right)
\end{equation}
where $I_\text{U} = 7.05$ g$\,$cm$^2$ is the moment of inertia of the cell from the bottom of the solid $^4$He sample to the top, $U = 3.43$ cm is the total length of this part of the cell, $\bar{\rho}$ and $\bar{\mu}$ are the weighted averages of the density and shear modulus of the cell$ + ^4$He system. Specifically for our cell, since $I_\text{U} \gg I_\text{He}$, $\bar{\rho} \approx \rho_\text{al}$ where $\rho_\text{al}$ = 2.7 g$\,$cm$^{-3}$ is the density of aluminum. $\bar{\mu}$ is given by considering the cross-section of the sample space. At radius $r < 0.635$ cm and $0.794$ cm $ < r <$ 0.921 cm, the cross-section is occupied with aluminum. Solid $^4$He only occupies the annular region 0.635 cm $ < r <$ 0.794 cm. Approximating the cell as a torsion rod with this given cross-section, we calculate that the fractional contribution of solid $^4$He to the total torsion constant is $0.48 \mu / \mu_\text{al}$ where $\mu_\text{al} = 2.9 \times 10^{11}$ dyn$\,$cm$^{-2}$ is the low temperature shear modulus of Al 6061 alloy. This suggests that $\bar{\mu} \approx \mu_\text{al} + 0.48 \mu$. 

As before, we vary $\mu$ from $1.5 \times 10^8$ dyn$\,$cm$^{-2}$ to $3 \times 10^8$ dyn$\,$cm$^{-2}$, obtaining a change in $I_\text{E}$ of $\Delta I_{\text{E}-} = 1.05 \times 10^{-6}$ g$\,$cm$^{2}$ for the low frequency mode and $\Delta I_{\text{E}+} = 6.01 \times 10^{-6}$ g$\,$cm$^{2}$ for the high frequency mode. Normalizing these values by the moment of inertia of solid $^4$He, $I_\text{He}$, the FPS is $4.23 \times 10^{-6}$ for the low frequency mode and $2.41 \times 10^{-5}$ for the high frequency mode. We note that this effect has the same $f^2$ dependence as the acceleration effect. Comparing the values of the FPS calculated here to those calculated for the acceleration effect, we see the finite shear modulus of the cylindrical cell wall alone enhances the acceleration effect by about 20$\%$.

\subsection{Torsion Rod Effect}
The effect on TO periods produced by the stiffening of solid $^4$He inside the fill-line drilled through the torsion rods is minimized by reducing the radius of the fill-line, $r_\text{fill}$. For our torsion rods which have outer radius $r_\text{rod}$ and shear modulus $\mu_\text{al}$, the FPS produced by a change in solid $^4$He shear modulus $\Delta \mu$ is \cite{Beamish_TorsionRodEffect}, in the limit of $r_\text{fill} \ll r_\text{rod}$,
\begin{equation}
\frac{\Delta P_\pm}{\Delta P_{\text{F}\pm}} \approx \frac{P_\pm}{2 \Delta P_{\text{F}\pm}} \frac{\Delta \mu}{\mu_\text{al}} \left( \frac{r_\text{fill}}{r_\text{rod}} \right)^4
\end{equation}
For our TO, $r_\text{fill} / r_\text{rod} = 0.067$. The measured mass-loading sensitivities are $\Delta P_{\text{F}-} = 2.75$ $\mu$s and $\Delta P_{\text{F}+} = 1.69$ $\mu$s. Assuming the shear modulus of solid $^4$He changes by 100$\%$ so that $\Delta \mu = 1.5 \times 10^8$ dyn$\,$cm$^{-2}$, the FPS is $2.94 \times 10^{-6}$ for the low frequency mode and $2.00 \times 10^{-6}$ for the high frequency mode. Since these estimates are two orders of magnitude smaller than the measured signals, we conclude that the solid $^4$He inside the torsion rods forms negligible contribution to the observed FPS.

\subsection{Maris Effect}
A subtle effect was discussed by H.~J.~Maris \cite{Maris_Effect} where the solid $^4$He sample inside the TO cell modifies the torsion constant of the torsion rod by exerting a counter-stress on the deformed cell wall separating solid $^4$He from the torsion rod. This ``Maris effect'' is much more significant for solid $^4$He samples with cylindrical geometries than for those with annular geometries. To estimate the size of such an effect in our annular cell, we replace the inner Al cylinder of the cell by solid $^4$He, so that the solid $^4$He sample is a cylinder with height $L = 3.28$ cm and radius $r_\text{o} = 0.794$ cm. The result from this simplification is an upper bound on the magnitudes of the signals if the actual geometry is used.

Following the approach outlined in Ref.~\cite{Maris_Effect}, we calculate that for a change of $\Delta \mu$ in solid $^4$He shear modulus, the corresponding change in the torsion constant $k_\text{c}$ of the cell $\Delta k_\text{c}$ is
\begin{equation}
\Delta k_\text{c} = 3.2 \times 10^{-9} (\Delta \mu / \mu) k_\text{c}
\end{equation}
It is noteworthy that the Maris effect only affects the torsion constant of the cell, whereas the torsion rod effect affects the torsion constants of both the cell and the dummy oscillator. Consequently, the frequency dependences of the two effects are different.

The induced period shifts $\Delta P_\pm$ are obtained by increasing $k_\text{c}$ by an amount $\Delta k_\text{c}$ and calculating the decrease in the TO periods $P_\pm = 2 \pi \left[ \tfrac{I_\text{c} k_\text{c} + I_\text{c} k_\text{d} +I_\text{d} k_\text{c}}{2 I_\text{c} I_\text{d}} \left( 1 \pm \sqrt{1 - \tfrac{4 I_\text{c} I_\text{d} k_\text{c} k_\text{d}}{(I_\text{c} k_\text{c} + I_\text{c} k_\text{d} +I_\text{d} k_\text{c})^2}} \right) \right]^{- 1/2}$. For a 100$\%$ change in $\mu$, the period shifts are $\Delta P_- = 2.5 \times 10^{-4}$ ns and $\Delta P_+ = 9.0 \times 10^{-4}$ ns. These values give FPS of $9.1 \times 10^{-8}$ for the low frequency mode and $5.3 \times 10^{-7}$ for the high frequency mode, both being more than two orders of magnitude smaller than the observed FPS.

\subsection{Dissipation of Solid $^4$He}
The additional dissipation introduced by the solid $^4$He in our experiment is very small. The increase in $1/Q$ where $Q$ is the mechanical qualify factor of the TO after the cell is filled with solid $^4$He has a peak value of only $5 \times 10^{-8}$ around a temperature of 0.1 K for both modes. The shifts in resonance periods associated with such changes in dissipation are $\Delta P_- = 3.87 \times 10^{-9}$ ns and $\Delta P_+ = 1.62 \times 10^{-9}$ ns. These values correspond to FPS of $1.41 \times 10^{-12}$ for the low frequency mode and $0.96 \times 10^{-12}$, eight orders of magnitude smaller than the FPS measured in the experiments.

\section{FEM Computations}

\begin{figure}
\centering
\includegraphics[width=\columnwidth]{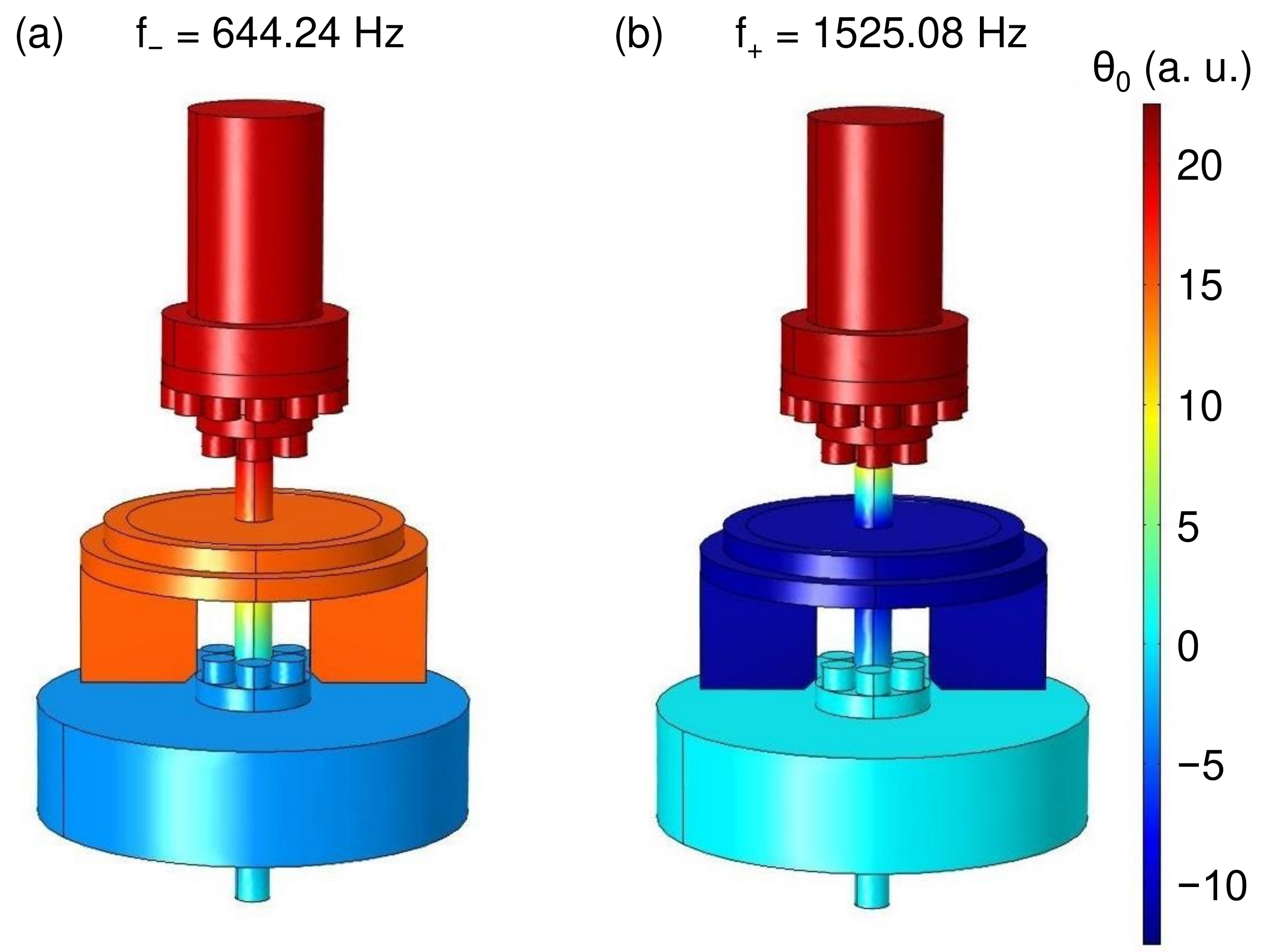}
\caption{(Color Online) FEM simulations of the amplitude of angular displacements $\theta_0$ and resonance frequencies $f_\pm$ of the two resonance modes of the annular TO used in the Letter. All dimensions used in building the FEM model are the measured values of the actual apparatus. $\theta_0$ is plotted in arbitrary units. The bottom oscillator is the vibration isolator.}
\label{fig:1}
\end{figure}

\begin{figure}
\centering
\includegraphics[width=\columnwidth]{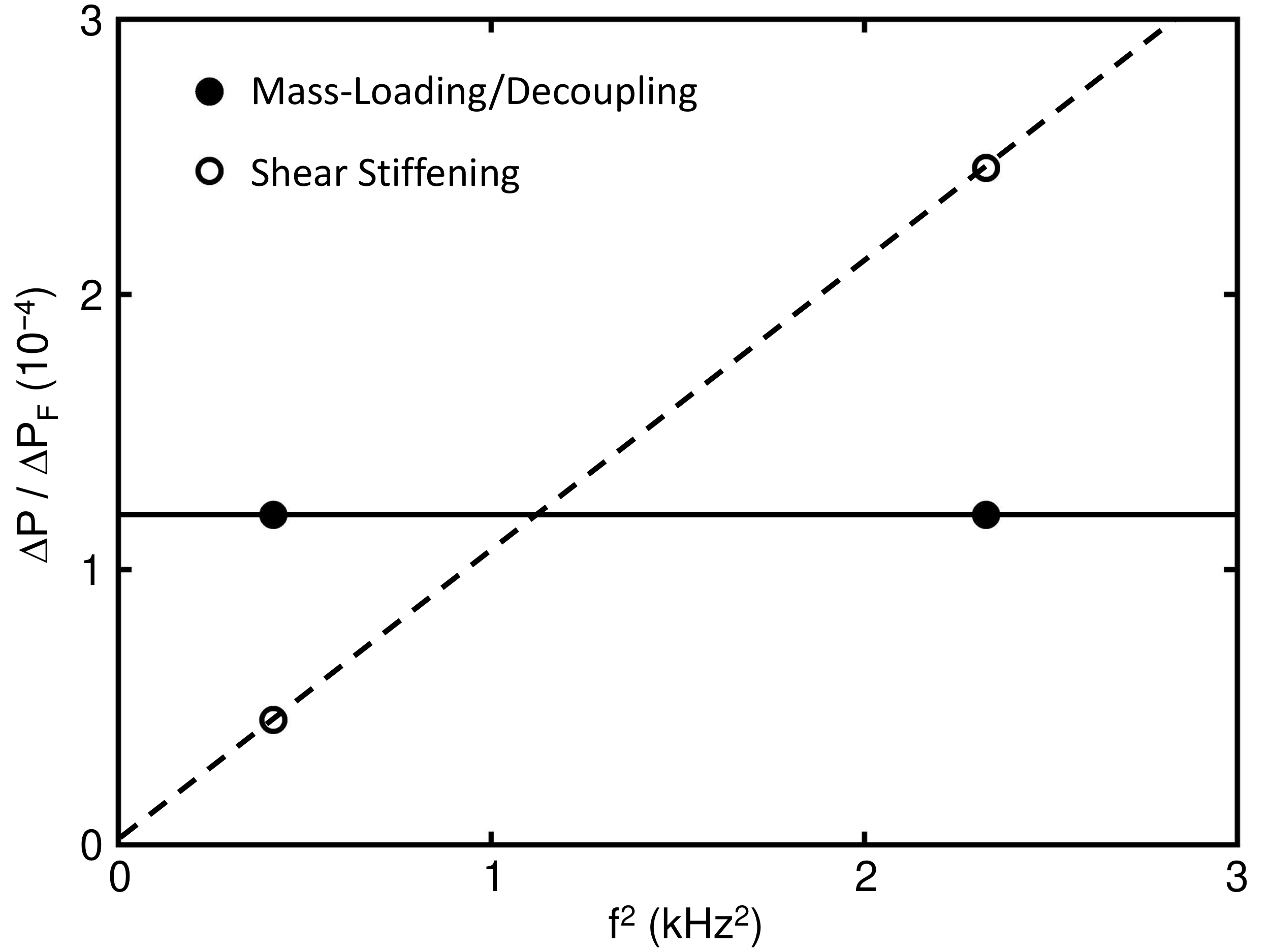}
\caption{FEM computations of FPS's ($\Delta P / \Delta P_F$) for the annular TO, based on a fractional change of $1.2 \times 10^{-4}$ in the density of solid $^4$He sample alone, and based on a 100$\%$ increase in the shear modulus of solid $^4$He sample alone.}
\label{fig:2}
\end{figure}

FEM calculations of TO resonance periods are performed with commercial software package COMSOL Multiphysics (structural mechanics module, COMSOL Multiphysics v4.3b, COMSOL Inc., 2013). A mesh consisting of 20745 domain elements is created, and the program solves for the eigen-frequencies $\omega$ of the Navier-Cauchy equation for the amplitude of the displacement field, $\vec{u}$, of the entire TO:
\begin{equation}
- \rho \omega^2 \vec{u} - \nabla \cdot \bar{\bar\sigma} = 0
\end{equation}
where $\bar{\bar\sigma}$ is the stress tensor. A plot for the amplitude of angular displacement, $\theta_0$, at the two resonance modes is shown in Figure.~\ref{fig:1}. We have included the vibration isolator in the model for added precision. It can be seen that the values of $\theta_0$ have the same sign at the cell and the dummy oscillator for the low frequency mode, but opposite signs at the high frequency mode. Hence the two modes have shapes expected from analytical calculations, where the cell and the dummy oscillator rotate in-phase at the low frequency mode and out-of-phase at the high frequency mode. From the values of $\theta_0$, we also extract the scale factors $D_\pm$ relating the angular velocity of the dummy oscillator $\dot{\theta}_\text{d}$ to that of the cell $\dot{\theta}_\text{c}$, $\dot{\theta}_\text{c} = D_\pm \dot{\theta}_\text{d}$, which turns out to be $D_- = 1.40$ and $D_+ = -1.65$. From Figure.~\ref{fig:1}, one can see the eigen-frequencies calculated by the program match those measured experimentally very closely. We also check for the accuracy of the model by varying the density of solid helium $\rho$ from 0 to 0.2 g$\,$cm$^{-3}$. The calculated shifts in resonance periods are 2.76 $\mu$s and 1.72 $\mu$s for the low and high frequency modes respectively, again matching the experimental values $\Delta P_{\text{F}-}$ and $\Delta P_\text{F+}$.

To study the effect of changing solid $^4$He shear modulus $\mu$, we shift the value of $\mu$ from $1.5 \times 10^8$ dyn$\,$cm$^{-2}$ to $3 \times 10^8$ dyn$\,$cm$^{-2}$ and calculate the shifts in periods at the two modes, $\Delta P_\pm$. Normalizing these shifts by the mass-loading values $\Delta P_{\text{F} \pm}$, the calculated FPS are $0.454 \times 10^{-4}$ for the low frequency mode and $2.46 \times 10^{-4}$ for the high frequency mode. These values are about twice those calculated with the analytical approach and likely to be quantitatively accurate, since the finite shear modulus of the entire TO is taken into account. They also suggest that a 50$\%$ change in the shear modulus is sufficient to account for the elastic contribution to the FPS observed in this experiment.

An important message imparted by the FEM simulation is the frequency dependence of the signals produced by the changing shear modulus of solid $^4$He. The ratio of the computed FPS at the two modes is $2.46 / 0.454 = (f_+ / f_-)^{1.96}$. The exponent of 1.96 is in excellent agreement with the analytical prediction of 2, suggesting that changing solid $^4$He shear modulus indeed produces a FPS that is proportional to $f^2$ for the cell presented in the Letter. To visualize such a frequency dependence, we plot the computed FPS values based on changes in $\mu$ in Figure.~\ref{fig:2} as a function of $f^2$. The values are seen to extrapolate to $<10^{-5}$ in the zero-frequency limit. In contrast, the value of FPS based on a supersolid fraction of $1.2 \times 10^{-4}$ is seen to be independent of frequency. Comparing the FEM computation to data presented in Figure.~5 of the Letter, we see that the experimental FPS values can only arise from a combination of both effects.

\bibliographystyle{apsrev}
\bibliography{references}